\documentclass[twocolumn,preprintnumbers,amsmath,amssymb,showpacs]{revtex4}
\usepackage{epsfig}
\begin{document}
\setlength{\voffset}{1.0cm}
\title{The phase diagram of the massive Gross-Neveu model, revisited}
\preprint{FAU-TP3-05/5}
\author{Oliver Schnetz}
\author{Michael Thies}
\author{Konrad Urlichs}
\affiliation{Institut f\"ur Theoretische Physik III,
Universit\"at Erlangen-N\"urnberg, D-91058 Erlangen, Germany}
\date{\today}
\begin{abstract}
The massive Gross-Neveu model is solved in the large $N$ limit at finite temperature and chemical potential.
The phase diagram features a kink-antikink crystal phase which was missed in previous works. Translated into
the framework of condensed matter physics our results generalize the bipolaron lattice in non-degenerate
conducting polymers to finite temperature.
\end{abstract}
\maketitle
%################################################################################################
%                                                                                           SECTION 1
%################################################################################################
In its original form, the Gross-Neveu model \cite{1} is a relativistic, renormalizable quantum field theory of $N$ species
of self-interacting fermions in 1+1 dimensions with Lagrangian
\begin{equation}
{\cal L} = \sum_{i=1}^N \bar{\psi}^{(i)} ({\rm i}\gamma^{\mu}\partial_{\mu} - m_0)\psi^{(i)} + \frac{1}{2} g^2 \left(\sum_{i=1}^N 
\bar{\psi}^{(i)}\psi^{(i)}\right)^2.
\label{A1}
\end{equation}
The bare mass term $\sim m_0$ explicitly breaks the discrete chiral symmetry $\psi\to \gamma^5 \psi$ of the massless model.
As far as the phase diagram is concerned, the 't Hooft limit $N\to \infty$, $g^2 \sim 1/N$ is most interesting
since it allows one to bypass some of the limitations of low dimensions, while justifying a semi-classical approach.
Interest from the particle physics side stems from the fact that the simple Lagrangian (\ref{A1}) shares
non-trivial properties with strong interaction physics, notably asymptotic freedom,
dimensional transmutation, meson and baryon bound states, chiral symmetry breaking in the vacuum as well as its
restoration
at high temperature and density (for a review, see e.g.~\cite{2}). It also has played some role as a testing ground for
fermion
algorithms on the lattice \cite{3}.
Perhaps even more surprising and less widely appreciated is the fact that Gross-Neveu type models have enjoyed
considerable success in describing a variety of
quasi-one-dimensional condensed matter systems, ranging from the Peierls-Fr\"ohlich model \cite{4} over ferromagnetic
superconductors \cite{5} to conducting polymers, e.g. doped {\em trans}-polyacetylene \cite{6}. By way of example, the kink
and kink-antikink
baryons first derived in field theory in the chiral limit ($m_0=0$) \cite{7} have been important for understanding the role of solitons
and polarons in
electrical conductivity properties of doped polymers \cite{6}. Likewise, one can show that baryons in the massive Gross-Neveu
model
($m_0 \neq 0$) are closely related to polarons and bipolarons in polymers with non-degenerate ground states. e.g.
{\em cis}-polyacetylene
\cite{8,9}. Finally, Gross-Neveu models with finite $N$ have been found useful for describing electrons in carbon nanotubes
and fullerenes \cite{10}.

The claim that a relativistic theory is relevant for condensed matter physics is at first sight very provocative \cite{11}. A closer
inspection shows that a continuum approximation to a discrete system, a (nearly) half-filled band and a linearized dispersion
relation of the electrons at the Fermi surface are the crucial ingredients leading to a
Dirac-type theory. The Fermi velocity plays the role of the velocity of light and the band width the
 role of the UV cutoff.
In some cases the correspondence is so close that results can be taken over from one field into the other.
This was illustrated in \cite{17} where we borrowed results from the theory of non-degenerate conducting
polymers, in particular the bipolaron lattice, for solving the zero temperature limit of the massive Gross-Neveu model.

The Lagrangian (\ref{A1}) has two bare parameters, $g^2$ and $m_0$. In the process
of regularization and renormalization all observables can be expressed in terms of two physical parameters $m$ and $\gamma$.
The relation to the bare quantities and the UV cutoff $\Lambda$ is given by the vacuum gap equation
\begin{equation}
\frac{\pi}{Ng^2} = \gamma + \ln \frac{\Lambda}{m}, \qquad  \gamma:= \frac{\pi}{Ng^2}\frac{m_0}{m}.
\label{A2}
\end{equation}
Whereas $m$ merely provides the overall mass scale and can be set equal to 1, the renormalized fermion mass ratio $\gamma$
(called ``confinement parameter" in condensed matter physics) parametrizes different physical theories. It measures the amount
of explicit chiral symmetry breaking and vanishes in the massless ($m_0=0$) case. The phase diagram of the model (\ref{A1})
was proposed in 1995 \cite{12}, but recent findings showed that the result is flawed. A related problem in the massless
Gross-Neveu model was pointed out and subsequently cured by us in previous works \cite{13,14,15}. We will recover these findings
below as special case $\gamma=0$ of the present work.

Let us first recall the original analysis \cite{12} which is still valid in a certain region of the phase diagram.
In the large $N$ limit, a saddle point approximation to the functional integral or, equivalently, a (thermal) Hartree-Fock
approach becomes exact. One finds that the fermions condense to form a scalar potential $S(x)$.
The authors proceed along similar lines as the 1985 work at $\gamma=0$ \cite{16} and tacitly assume that $S$ is spatially constant
leading to a ($\mu,T$)-dependent dynamical fermion mass $M$. This effective mass is determined via self-consistency by the thermal
expectation value of the fermion condensate,
\begin{equation}
M=m_0  -Ng^2 \langle\bar{\psi}\psi\rangle_{\rm th}.
\label{A3}
\end{equation}
For a translationally invariant system, this in turn is tantamount to minimizing the renormalized
grand canonical potential density
\begin{eqnarray}
\Psi &=& \frac{M^2}{2\pi}\left( \ln M-\frac{1}{2}\right)+\gamma\left(\frac{M^2}{2\pi}-\frac{M}{\pi}\right)
\label{A4} \\
&- & \frac{1}{\beta \pi} \int_0^{\infty} {\rm d}q \ln \left[ \left( 1+ {\rm e}^{-\beta(E-\mu)}\right) \left( 1+
 {\rm e}^{-\beta(E+\mu)} \right)\right]
\nonumber
\end{eqnarray}
($m=1, E=\sqrt{q^2+M^2}$) with respect to $M$. Depending on the parameters ($\mu,T,\gamma$), there may be
one or two local minima with the possibility of a first order phase transition. The phase diagram depending on
$\mu,T,\gamma$ is plotted in Fig.~1.
\begin{figure}[h]
%\begin{center}
\epsfig{file=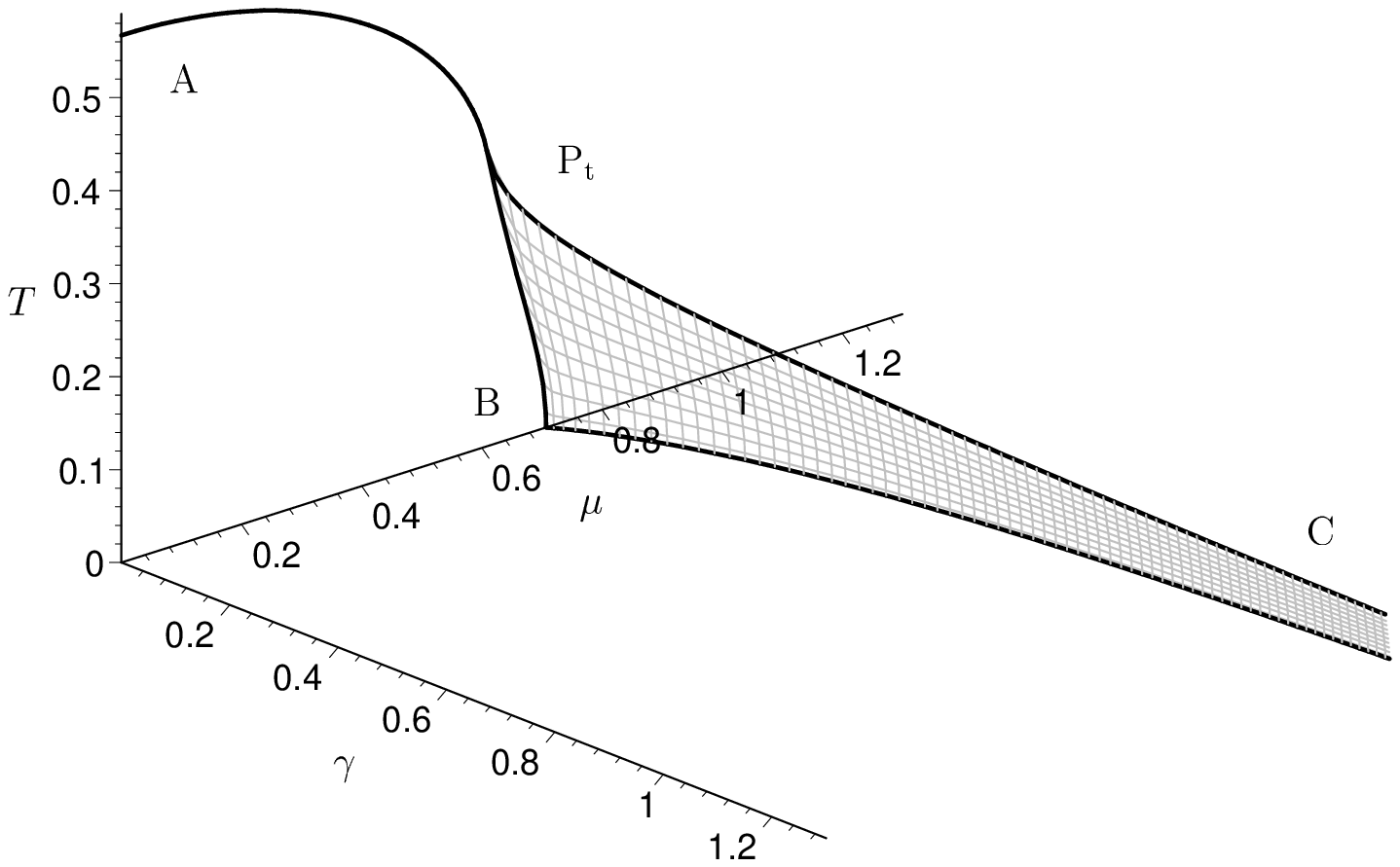,width=5.0cm,angle=270}
\caption{Phase diagram of the massive Gross-Neveu model as a function of $\gamma,\mu,T$, (incorrectly) assuming unbroken translational
invariance. The various phase boundaries are explained in the main text.}
%\end{center}
\end{figure}
In the chiral limit ($\gamma=0$), the massive phase
at low $(\mu,T)$ is separated from the chirally restored, massless phase at high $(\mu,T)$ by a critical line (AP$_{\rm t}$B)
\cite{16}. The upper part  of this line (AP$_{\rm t}$) is 2nd order, the lower part (P$_{\rm t}$B) first order, with a tricritical
point P$_{\rm t}$ separating the two. As the parameter $\gamma$ is switched on,
the 2nd order line disappears in favour of a cross-over where the fermion mass changes rapidly, but smoothly.
The first order line on the other hand survives, ending at a critical point. If plotted against $\gamma$, these
critical points lie on the third curve (P$_{\rm t}$C) emanating from the tricritical point shown in Fig.~1. For $\gamma>0$ the effective
fermion mass $M$ is different from zero everywhere. If one crosses the shaded critical ``sheet" in Fig.~1, the mass
changes discontinuously, dropping with increasing chemical potential. More details on the thermodynamic implications of such
a phase diagram are given in \cite{12}.

In the meantime, evidence has accumulated that this picture cannot be correct. This has so far only been
clarified in the two coordinate planes $\gamma=0$ \cite{15} and $T=0$ \cite{17}
of Fig.~1, and we propose to complete the phase diagram here. Briefly, what has been missing is a proper treatment of the
baryons which can give rise to a kink-antikink crystal phase in certain regions of temperature and chemical potential.
Actually, this phenomenon has been known in condensed matter physics for a long time: It reflects the Peierls instability,
a general phenomenon in one-dimensional systems. In order to lower the energy or the
thermodynamic potential, a periodic modulation of the lattice is induced, thereby opening a gap at  the Fermi surface.
In the relativistic Gross-Neveu model it turns out that the scalar condensate $\langle \bar{\psi}\psi \rangle$
(the $\sigma$-field) plays the role of the ion displacement.

To incorporate this kind of physics into the Gross-Neveu model, one has to solve the Dirac equation self-consistently
(including the
negative energy sea) with a spatially varying, periodic scalar potential $S(x)$, in general a highly non-trivial task
even in one space dimension.
Fortunately, in the special cases considered so far (namely $\gamma=0$ and $T=0$), this has turned out to be exactly feasible
due to a number of lucky circumstances. The self-consistent potentials are such that the Dirac equation
 can be
mapped onto a 2nd order Schr\"odinger type equation with the Lam\'e potential of order 1.
The single particle eigenspinors are known in terms of theta functions. This enables one to perform an exact
calculation of both the ground state and the phase diagram, including an analytical proof of self-consistency.
At $\gamma=0$, such a calculation was done independently in condensed matter physics \cite{4} and
 relativistic field theory \cite{15}.
During the last two decades, interest in non-degenerate conducting polymers has fostered a lot of theoretical work on the
 polymer side.
As a result, analytical bipolaron lattice solutions at zero temperature were found by several groups \cite{18,19,20,21,22}.
We have recently shown that the $x$-dependent gap parameter of the bipolaron lattice is adequate for solving the $T=0$,
massive Gross-Neveu model \cite{17}.

We have extended this calculation to finite temperature and again found exact self-consistency. We can therefore now present
a revised phase diagram in the 3-dimensional ($\mu,T,\gamma$)-space which supersedes the old, translationally
invariant phase diagram of Fig.~1. Whereas the solution demands some mathematics (similar but more complex than the $\gamma=0$
solution presented in \cite{15}) the outcome can be visualized in a single graph. This made us decide to present in this Letter
our main results without derivation. The detailed formalism and further results
on the revised thermodynamics of the massive Gross-Neveu model will be given elsewhere.

The crucial observation is that the scalar potential at $T\neq 0$ has the same general functional form as at $T=0$. It can be
written in terms of three parameters $A, b, \kappa$ and Jacobi elliptic functions of the modulus $\kappa$ (suppressed here) as
\begin{equation}
S(x) = A\left( \kappa^2 {\rm sn}\, b\, {\rm sn}\, Ax \, {\rm sn} (Ax +b) + \frac{{\rm cn} \,b\, {\rm dn} \,b}{{\rm sn} \,b}\right).
\label{A6}
\end{equation}
This form equals the more complicated expression employed in the literature on the bipolaron lattice and
in our $T=0$ paper \cite{17}. As mentioned above,
the Dirac equation with potential (\ref{A6}) is equivalent to the single gap
Lam\'e equation. Using the known spectrum and density of states of the Lam\'e potential,
one can construct the grand canonical potential and minimize it with respect to ($A,b,\kappa$).
Due to the specific structure of the equations obtained, it is possible to reduce the solution to basically
solving a transcendental equation in one variable containing a one-dimensional numerical integration.
The proof of the self-consistency condition which replaces Eq.~(\ref{A3}),
\begin{equation}
S(x)=m_0  -Ng^2 \langle\bar{\psi}\psi\rangle_{\rm th},
\label{A7}
\end{equation}
establishes that the minimum of the grand canonical potential is an exact solution of the large $N$ Gross-Neveu model.
Comparison with the translationally invariant solution shows that the crystal is thermodynamically favourable whenever it exists.

We now turn to the results, focussing on the phase boundaries as depicted in Fig.~2.
\begin{widetext}
\begin{figure}[h]
\begin{center}
\epsfig{file=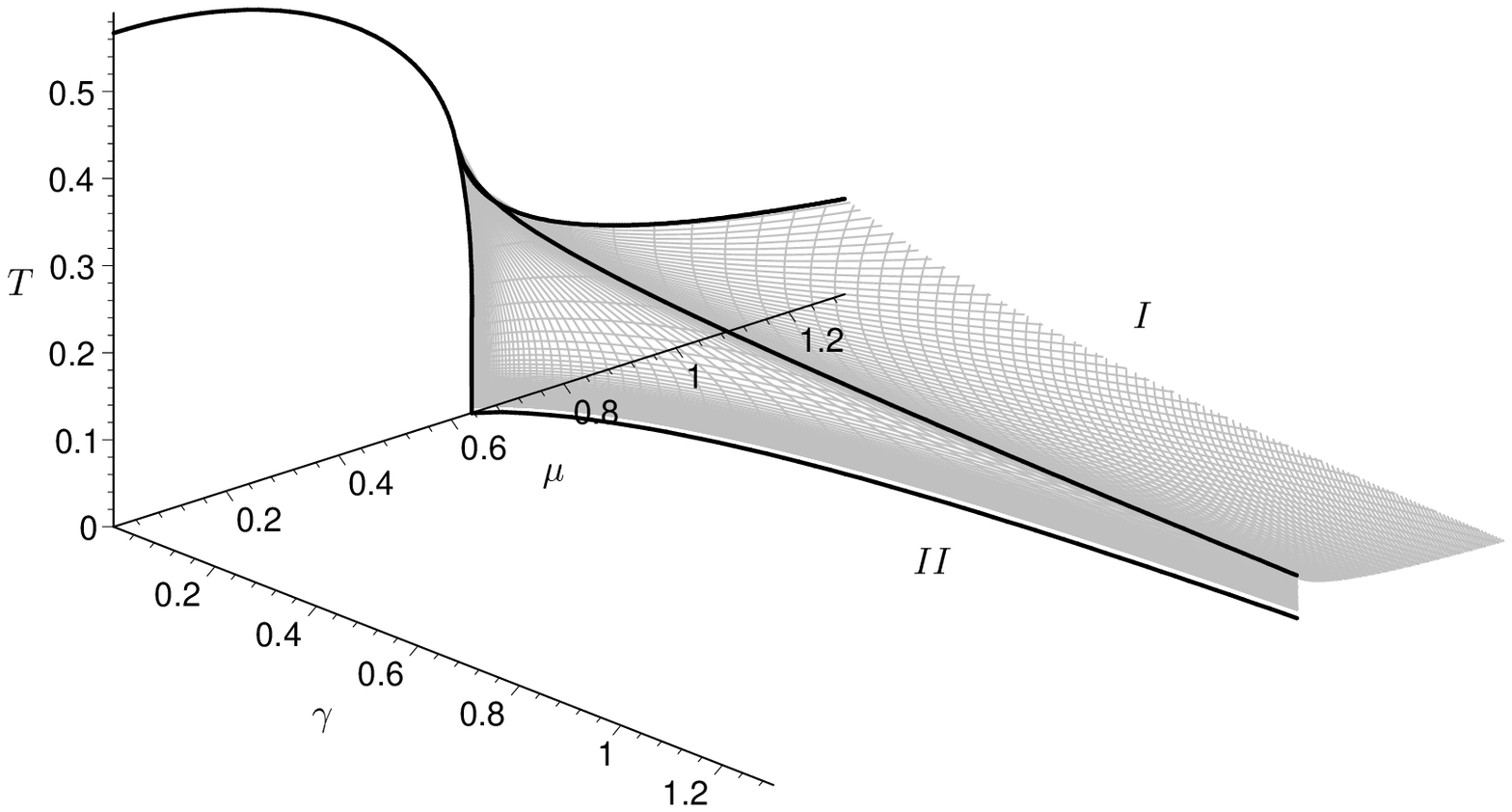,width=8cm,angle=270}
\caption{Revised phase diagram of the massive Gross-Neveu model. The shaded surfaces $I,II$ separate the kink-antikink
crystal from a massive Fermi gas and correspond to 2nd order phase transitions.}
\end{center}
\end{figure}
\end{widetext}
In the $\gamma=0$ plane, the result for the revised phase diagram including the kink-antikink crystal is known
\cite{15}: The 2nd order line (AP$_{\rm t}$) of the old phase diagram is unaffected. The first order line
(P$_{\rm t}$B) of Fig.~1 is replaced by two 2nd order lines delimiting the kink-antikink crystal phase. The
tricritical point is turned into another kind of multicritical point at the same ($\mu,T$) values. As
we turn on $\gamma$, the second order line separating massive ($M>0$) and massless ($M=0$) phases disappears
as a consequence of the explicit breaking of chiral symmetry. The crystal phase survives at all values of $\gamma$, but is confined
to decreasing temperatures with increasing $\gamma$. For fixed $\gamma$, it is bounded by two 2nd order lines joining in a
cusp. The cusp coincides with the critical point of the old phase diagram but has a significantly different
character. The crystal phase exists and is thermodynamically stable inside the tent-like structure formed out of two sheets
denoted as $I$ and $II$. These sheets are defined
by $\kappa=0$ ($I$) and $\kappa=1$ ($II$), respectively. The line where they join corresponds to $b=0$
and coincides with line (P${\rm _t}$C) in Fig.~1. The baseline of sheet $II$ in the ($\mu,\gamma$)-plane
has a simple physical interpretation: It reflects the $\gamma$-dependence of the baryon mass in the massive Gross-Neveu
model, or equivalently the critical chemical potential at $T=0$.
\begin{figure}[h]
\begin{center}
\epsfig{file=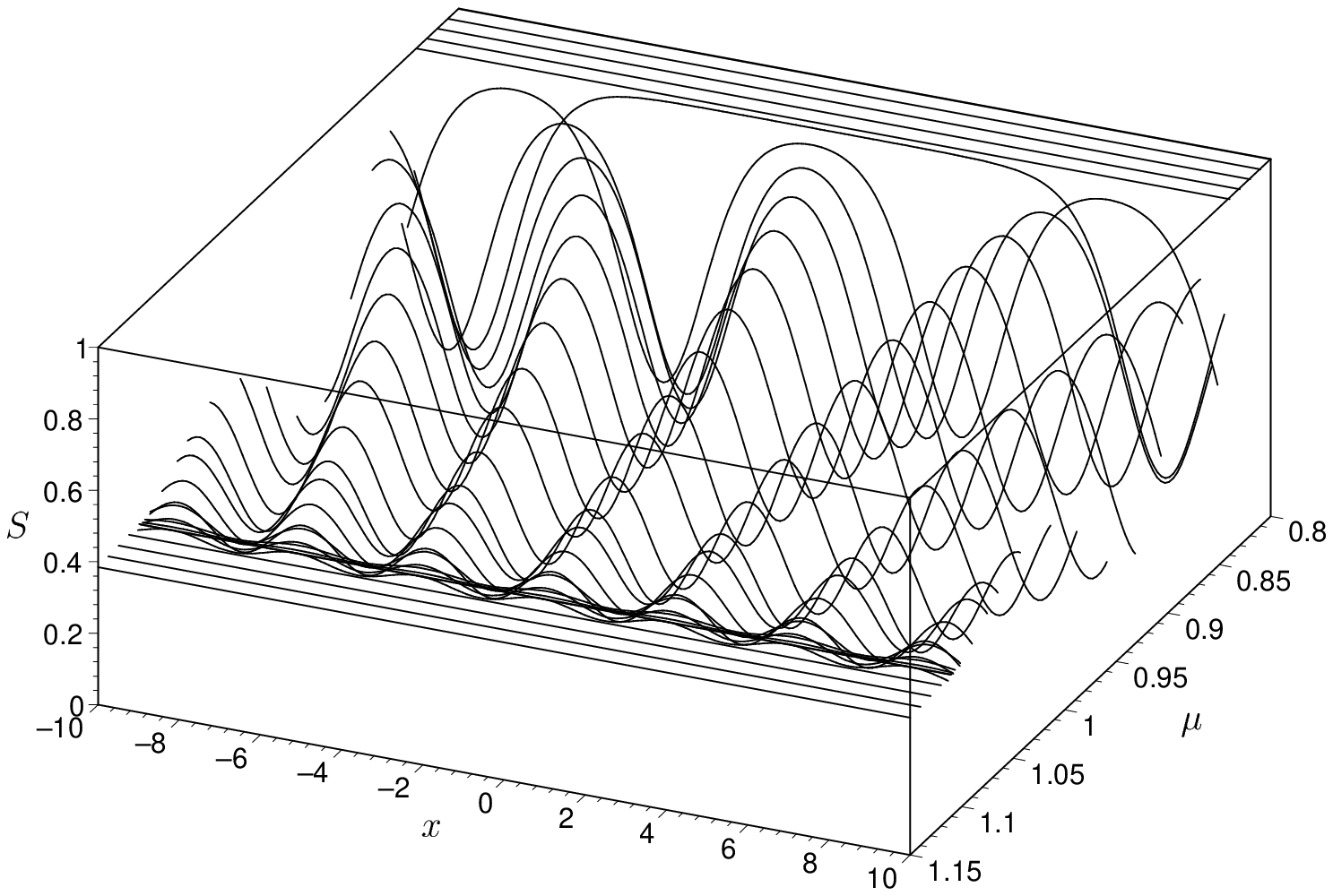,width=5cm,angle=270}
\caption{Evolution of self-consistent $S(x)$ as one crosses the crystal region on a straight path. As $\mu$ decreases from
1.1 to 0.85, the elliptic modulus $\kappa$ varies between 0 and 1.}
\end{center}
\end{figure}
The chiral limit $\gamma \to 0$ can be identified with $b \to {\bf K}(\kappa)$. In this way, one can recover
the simpler 2-parameter ansatz for the self-consistent potential used in \cite{15} from the 3-parameter
ansatz Eq.~(\ref{A6}).

In order to further illustrate the nature of the phase transitions between inhomogeneous and homogeneous
phases, let us remark that outside the tent the scalar potential is $x$-independent, i.e. a dynamical mass.
If one crosses the tent from side $I$ to side $II$ on some curve, the parameter $\kappa$ varies continuously from 0 to 1.
By way of example consider the straight line $\gamma=0.5, T=0.05 \mu$ which pierces sheet $I$ at $\mu=1.1$
and sheet $II$ at $\mu=0.85$. Just outside of the crystal phase, the dynamical mass values we find are
0.4 ($I$) and 1.0 ($II$).
According to the old phase diagram, the mass jumps suddenly from 0.54 to 1.0 upon crossing
the first order sheet at $\mu=0.87$. By contrast, Fig.~3 shows the continuous evolution of $S(x)$ in the exact calculation.
At $\kappa=0$, an instability occurs with respect to oscillations of a finite wavenumber. As $\kappa$ increases,
the amplitude of these oscillations become larger whereas the period first grows rather modestly.
When $\kappa$ approaches the value 1 the period grows rapidly and, save for small dents representing widely spaced baryons,
$S(x)$ reaches a constant value
connecting to the translationally invariant phase. At $\kappa=1$ the system is instable against single baryon formation.
This subtle interpolation of $S(x)$ between two constants caused by the Peierls instability is only crudely modeled by the
translationally invariant scenario.

Concluding, we find it gratifying that the simple Lagrangian (\ref{A1}) gives rise to such a rich phase diagram.
We emphasize that we would not have been able to solve this problem without the works on the bipolaron lattice in
conducting polymers. We hope that our extension to finite temperature will in turn lead to new results in condensed matter
systems.
\\
\\
We should like to thank Wolfgang H\"ausler for helpful conversations on one-dimensional condensed
matter physics.

\end{document}